# ANALYSIS OF LITHOGRAPHY BASED APPROACHES IN DEVELOPMENT OF SEMICONDUCTORS

Jatin Chopra[1]

[1]Department of Computer Science, BITS Pilani Dubai, United Arab Emirates

## ABSTRACT

*The end of the 19th century brought about a change in the dynamics of computing by the development of the microprocessor. Huge bedroom size computers began being replaced by portable, smaller sized desktops. Today the world is dominated by silicon, which has circumscribed chip development for computers through microprocessors. Majority of the integrated circuits that are manufactured at present are developed using the concept of Lithography. This paper presents a detailed analysis of multiple Lithography methodologies as a means for advanced integrated circuit development. The study paper primarily restricts to examples in the context of Lithography, surveying the various existing techniques of Lithography in literature, examining feasible and efficient methods, highlighting the various pros and cons of each of them.*

## 1. INTRODUCTION

In 1950's chip manufacturers discovered that as the complexity of the circuits grew they faced new challenges [2]. One was while building a circuit, all the connections should be intact otherwise the current would be stopped within the circuit. Another problem was the speed complexity of the circuits. The circuits used in computers primarily depend upon speed. If the components used in computers are too large or the wires joining them are too long, then in that case the speed will become slow and get affected. In 1958 Jack Kilby established an answer to this problem. He made all the parts from the same block of material and added a layer of metal on top of it to connect them together. As a result no more of wires and components had to be assembled manually. The circuits could easily be made smaller and the manufacturing process could be computerized. Integrated circuits have come long way since the Jack Kilby's first model. His idea changed the future of microprocessor industry and is the key element behind our computer society.

Integrated circuit is considered as the advanced form of electrical circuit. Different electrical components including transistors, capacitors, diodes and capacitors make an electrical circuit. These components act as building blocks in an electrical manufacture gear. It can help us construct computer microprocessor or burglar alarm. The advent of integrated circuits (ICs) technology facilitated the placement of large number of transistors on a single chip, thereby playing an important role in solving problems of logic circuits by improving power consumption and speed performance.

Today most advanced circuits can be developed with billions of transistors embedded in them [5]. Every year as the technology is advancing; the chips are becoming more powerful and cheaper.

To build smaller and smaller chips, manufacturers have improved the accuracy of laser and lenses by reducing the wavelength of the light striking the wafer.

The method behind the construction of current ICs involves shining light through a mask onto a photoresist-coated loafer and subsequently removing the exposed areas. This process for manufacturing components of an IC is termed as Lithography. In order to achieve the circuitry pattern on to the loafer, several techniques of lithography can be applied. This paper will outline the most popular methods of Lithography that are employed in the semi-conductor industry which at a very high level can be categorized as [1] Electron-Beam Lithography, X-Ray Lithography and Optical Lithography.

*Structure of Paper*

The remainder of the paper is organized as follows: In Section 2, the objective of this paper is presented. In Section 3, the various acronyms used in this paper have been given. The fundamentals of Lithography comprising of the procedure and merits are presented in Section 4. Section 5 deals with detailed analysis of the various approaches for chip manufacture using Lithography. The paper is concluded in Section 6.

## 2. OBJECTIVE

The present work is intended to meet the following objectives:

1. Summarize the process of chip manufacture using Lithography, along with merits and demerits.
2. Survey the existing Lithography tools that are employed for the development of ICs.
3. Present a detailed analysis on Extreme Ultraviolet Lithography (eUV) as a promising technology for chip making.

## 3. ACRONYMS

1. AMD    : Advanced Micro Devices.
2. ARF    : Argon Fluoride.
3. DUV    : Deep Ultraviolet Lithography.
4. EUV    : Extreme Ultraviolet Lithography.
5. HF     : Hydrogen Fluoride.
6. HSQ    : Hydrogen Silises Quioscane.
7. IBM    : International Business Machines.
8. IC     : Integrated Circuit.
9. KRF    : Krypton Fluoride.
10. PMMA : Poly Methyl Methacrylate.
11. PBS   : Poly Butane -1 - Sulphone.

## 4. OVERVIEW OF LITHOGRAPHY

The term Lithography [3] originates from Greek words *lithos* meaning "stones" and *graphia* meaning "writing". In today's semiconductor industry it is used to design 3-D images on a substrate. It is a process of imprinting a geometric pattern from a mask onto a thin layer of materials called resist. The technology plays an important role in the production of electronic device components such as integrated circuits. Figure 1 shows the classification of lithography.

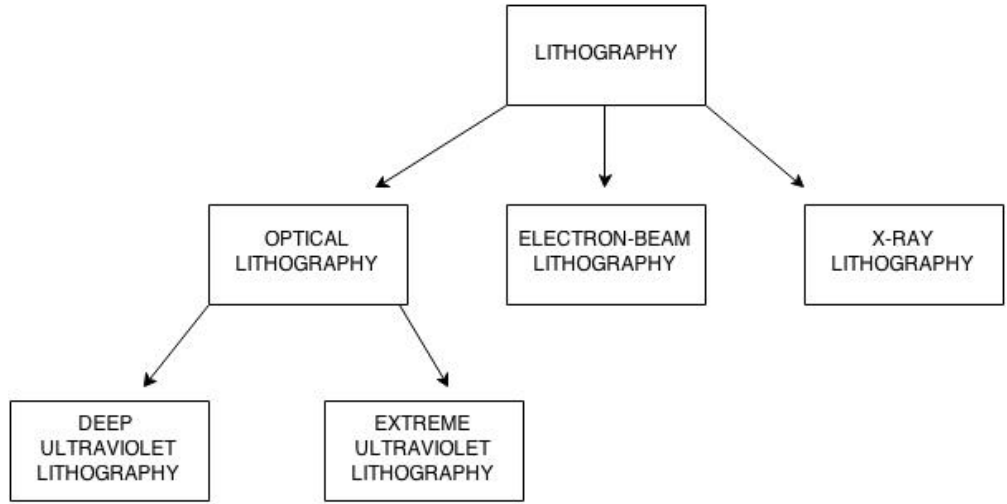

Figure 1.Classification of Lithography

In the following subsections, steps are discussed to create integrated circuits using lithography along with the terminology.

### 4.1. Terminology

1. *Wafer / Substrate*: The substance on which pattern of integrated circuits are developed by different steps of lithography
2. *Photoresist*: Is a photosensitive film placed on the top of the wafer
   There are two types of photoresist:-
   - Positive Photo resist: Becomes more soluble in developer solution when exposed to radiation.
   - Negative Photo resist: Becomes less soluble in developer solution when exposed to radiation.
3. *Pattern*: It is the image of the integrated circuit that has to be printed on the wafer

### 4.2. Procedure

The following are the steps involved in creation of integrated circuits using lithography (Figure2):

*Step 1*: Coat the wafer with poly silicon liquid and place the photo resist layer on top of it.

*Step 2*: Expose the photoresist-coated wafer to proper radiation. Project a light on the mask containing the pattern. Expose the light on the wafer with or without a mask. Some patterns on the mask are opaque and some are transparent to the radiation falling on them, so when the light falls on the mask only those regions which are transparent will allow the light to pass through it and fall on the wafer. Different resist have different properties. Properties of only those resist will change which are exposed to proper radiation and later become soft or hard.

*Step 3:* Developing stage involves two areas:
  Area 1: Photo resist becomes soft to radiation.
  - Place the resist in a developer solution for easy detachment. Only that resist will remain which is not yet exposed to radiation

- Dilute in Hydrogen Fluoride (HF) solution. The resist, which is guarding the oxide layer and the region that are now exposed to radiation, will get removed. This technique is termed as etching.
- Compare patterns on the wafer with the mask. The patterns that are opaque are same as present on the wafer.

Area 2: The pattern on the photoresist becomes hard. In that case the pattern on the wafer will be negative and will be compliment to the pattern on the mask, which is denoted as negative resist.

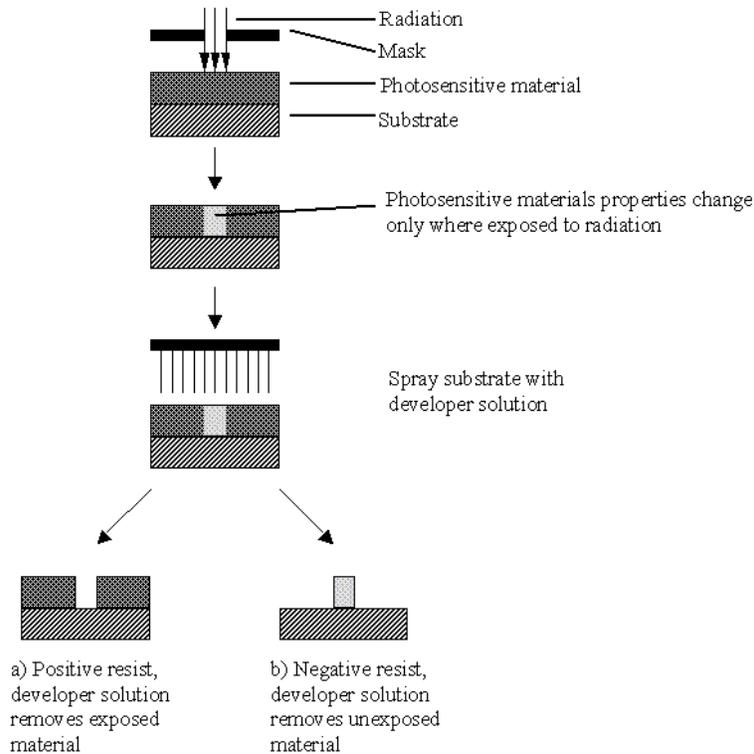

Figure 2. Procedure for Lithography

### 4.3. Precautions

The following are the precautions that need to be undertaken while creating integrated circuits using Lithography:

1. Dimensions of the pattern should be correct on the photo mask.
2. Transfer the pattern from the photo mask on the wafer carefully.
3. Photoresist should stick properly on the substrate because when photoresist is peeled off the pattern may get damage and become of no use.
4. The pattern fixed on the wafer should be aligned accurately as Lithography process can be carried out multiple times in a day.

## 5. LITHOGRAPHY METHODOLOGIES

### 5.1. Electron Beam Lithography

Electron Beam Lithography was discovered around 1960s in order to scan small objects with electrons and easily modify the patterns. The process of E-Beam Lithography involves bending of electrons using force as compared to conventional lithographic approach of bending light using mirrors [8]. This method is mask less where the substrate is directly written upon like a pen.

There are two types of forces involved

- Electromagnetic
- Electrostatic

Both of these forces are used to manufacture lenses that include electromagnetic and electrostatic lenses, which are used to focus light. In the following subsections the procedure used to create pattern on the substrate using E-Beam Lithography is discussed.

### 5.1.1 Procedure

The following are the steps involved in creation of integrated circuits using E-Beam Lithography

*Step 1:* E-Beam with a diameter of 0.01µ to 0.5µ is focused and scanned over the entire semiconductor substrate with a computer depending on the pattern that has to be created on the substrate. Since we use computer here, so it requires blanking (it controls the electron beam with the help of a computer and turns off and on accordingly) and deflection.

*Step 2:* The E-Beam is operated in two modes - one where the E-Beam Resist is eradiated changing its properties and other when off, the resist is not eradiated. The movement of the E-Beam is over a small region, which is referred as scanned field.

*Step 3:* Usually the size of the field is 1cm by 1cm whereas the size of the substrate is bigger in comparison. Once the E-Beam scans the substrate once, then the substrate gets moved, as a result the substrate is mounted over a movable X-Y table.

*Step 4:* Hence during the second time when the E-Beam is scanned over the unexposed regions of the substrate, patterns are created easily on it.

### 5.1.2 Writing Strategies

Following are the writing strategies for the creation of integrated circuits using E-Beam Lithography

1. *Raster Scan:* Scans the whole substrate pixel element by pixel element. It is a slow process and blanking is used when to write the pattern or not on the substrate accordingly. Spot size (smallest size that can be made on electrons and electron's numerical aperture. It defines the size ranging between 0.1 to 5 created over numerical aperture) plays an important role in writing since it gives direct tradeoff between throughput and resolution. The small spot size will provide better resolution whereas large spot size will provide higher throughput.
2. *Vector Scan*: Used only on the selected areas on the substrate where writing is to be performed and provides thinly dispersed and not dense patterns on the substrate.

International Journal of Computer Science & Information Technology (IJCSIT) Vol 6, No 6, December 2014

   3. *Variable shaped beam*: Projects a rectangle on the entire wafer at one time. These rectangles usually are of order 1 micron or half micron, so can easily adjust the width and height to get variety of shapes. This technique is used widely for the production of masks since it gives better throughput and resolution as compared to other techniques.

### 5.1.3. Advantages

- E-Beam Lithography is designed in such a manner that it delivers optimum performance in the field of research and development sector since it is automated and we can easily change the pattern when we want and is used in manufacturing masks.
- We obtain less throughput and high resolution.

### 5.1.4. Limitations

- *Dissolving Issues*: E-beam resists are organic materials which upon eradication of negative resist create cross linkage monomers and change the molecular weight at the same time. Making it heavy and difficult to dissolve in the developer solution.
- *Slow Process*: Since the diameter of the electron beam is very small only a small scanned area will be focused over and hence will take a longer time to write on the entire substrate [9].
- *Proximity Effect*: Different portions on the substrate have to be exposed for different amount of time. Figure shows two points A and B where A will receive the maximum electron beam when exposed and since point B is far from point A and will receive a very low amount of electron beam. This is termed as Proximity Effect. Since electron beam are used to write onto a small area of the substrate, multiple beams are being considered. Multiple beams are used altogether in parallel to write on the substrate and develop patterns easily. This helps to develop masks quickly but practically hasn't been demonstrated yet.

## 5.2. X-Ray Lithography

Wilhelm Conrad Roentgen discovered an unknown ray in 1895 (X-Ray) that can cause barium platinocyanide coated screen to glow. He won the first Nobel Prize in physics in 1901. In X-Ray Lithography X-Ray light is applied to generate the pattern on the substrate. Resolution is a function of wavelength and shorter the wavelength higher will be the resolution. Proximity printing is applied, where some gap is maintained between the substrate and the X-Ray mask. The technique is developed to the extent of batch processing. The method is simple to use since it requires no lenses. This technique originated as an advanced tool for next generation lithography for the semiconductor industry with batches of microprocessors successfully produced. Figure 3 below depicts how x-ray lithography is carried out.

66

International Journal of Computer Science & Information Technology (IJCSIT) Vol 6, No 6, December 2014

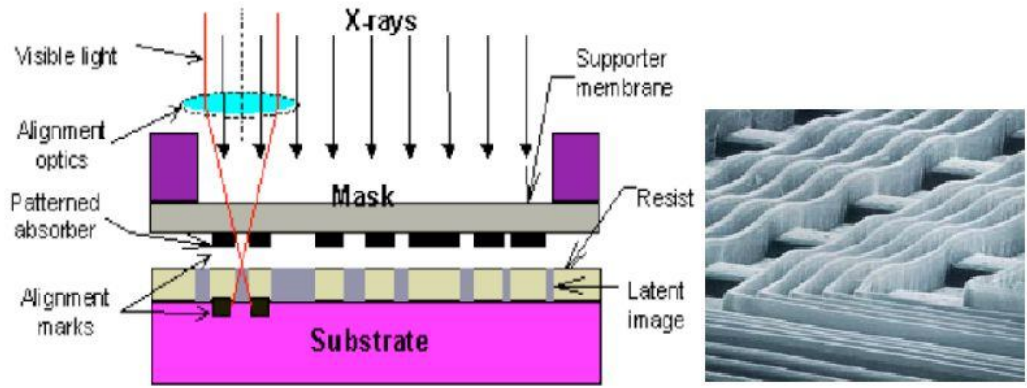

Figure 3. Procedure for X-Ray Lithography

### 5.2.1. Procedure

These are the steps that are to be followed during X-Ray Lithography:

*Step 1:* The source emits X-Ray light onto the X-Ray mask containing the pattern to be printed.
*Step 2:* X- Ray masks contain two important structures:
- *Absorber*: Contains the information to be imaged onto the resist. Since gold has a low atomic no and is a good absorbing material, it is applied on the top of the x-ray mask. High atomic number material absorbs X-Rays whereas low atomic number material transmits X-Rays.
- *Membrane*: X-Ray masks consist of a very thin membrane of low atomic number material carrying a high atomic number thick absorber pattern.

*Step 3:* X-Ray resist are of two types as shown in Figure 3:
- *Positive resist*: Which become soluble when exposed to X-Ray light in the developer solution. Example: PMMA and PBS. PMMA produces a resolution less than 0.1 whereas PBS produces a resolution of 0.5 [10]
- *Negative resist*: The portion, which is exposed to light, doesn't dissolve in the developer solution. Example: **COP** (An epoxy copolymer of glycidyl methacrylate and ethyl acrylate) which produces a resolution of 1.0.

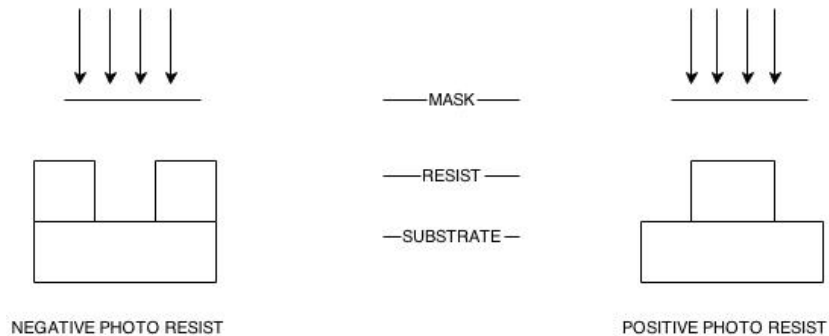

Figure 4.Types of X-Ray Resist

### 5.2.2. Advantages

- Reduces the limit of diffraction since shorter wavelength of light is used.
- It can produce feature size as small as 20nm.
- It produces high resolution and less through put.

- There is no proximity effect since secondary electrons don't have high energy
- X-Ray light is not absorbed by dirt and even if there is dirt present on the mask, the pattern won't get affected

### 5.2.3. Limitations

- It takes long time for the resist to get exposed.
- It is difficult and costly to manufacture X-Ray masks.

## 5.3. Optical Lithography

Optical Lithography is a process of transferring the pattern from the photographic mask onto the wafer. It is used in mass-production of ICs with minimum feature size of 250-350 nm. There are three ways of printing in optical Lithography onto the wafer:

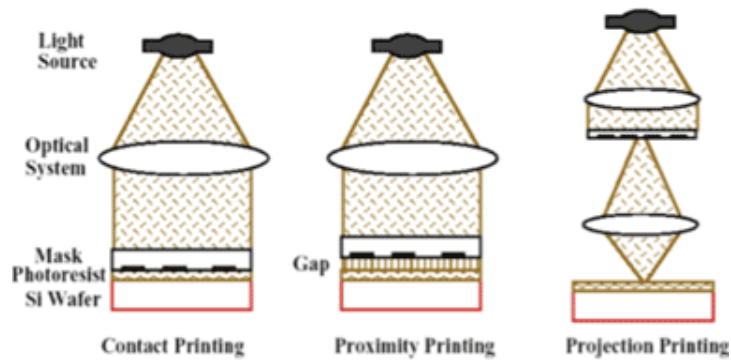

Figure 5. Types of Printing in Optical Lithography

1. *Contact Printing*: In Contact Printing the wafer and the mask come in close contact with each other, which makes the life of mask very low but creates a high resolution. It makes use of u-v light of wavelength 300-400nm. It operates in the Fresnel Diffraction pattern (used to calculate the diffraction pattern created by waves passing through an aperture or around an object). The tools used here are optical system, wafer, photoresist film and mask. When the light passes through the optical system, it makes the light incident on the glass pate. The silicon wafer is coated with photoresist film.

#### 5.3.1.1. Advantages

- Contact printing provides good resolution of images of about 1 to 0.5µm.
- It is easy to operate and maintain the equipment.

#### 5.3.1.2. Disadvantages

- Masks may get damaged due to contact with the wafer for a long time, so this technique is only limited to research labs or applications that can tolerate the defects.

2. *Proximity Printing*: In proximity printing a gap is maintained between the mask and the wafer, which makes the life of the mask better but does not provide great resolution. This can help in minimizing the mask getting damaged but still dust particles may be present and can damage the photo mask. It operates in the Fresnel diffraction pattern (used to calculate the diffraction pattern created by waves passing through an aperture or around an object) where resolution is proportional to $\sqrt{\lambda g}$ where $\lambda$ is the wavelength of light exposed and g is the gap between the mask and the wafer, since g is less here in comparison to contact printing it leads to poor resolution.

#### 5.3.2.1. Advantages

- Mask are less defect prone than in pure (hard) contact.

#### 5.3.2.2. Disadvantages

- It is hard to align and will lose some resolution of images due to the small gap between the mask and the wafer resulting in diffraction at feature edges.

3. *Projection Printing*: In projection printing the distance between the mask and the wafer isn't constrained because the pattern on the photo mask is projected directly on the resist coated wafer and there is no chance of the mask getting damaged as the mask and the wafer are placed far away from each other. It operates in the Fraunhofer diffraction pattern (used to model the diffraction of waves when the diffraction pattern is viewed at a long distance from the bending object). In order to obtain an image of high resolution, only a small amount of pattern is projected and for a defect free lens system Rayleigh limit formula is applied

$0.6 \lambda : NA$ where $\lambda$ is the wavelength used and NA is the numerical aperture of the lens system used.

#### 5.3.3.1. Advantages

- Provides high resolution of images.
- There is no damage to mask.

#### 5.3.3.2. Disadvantages

- Complex equipment requires high maintenance
- Provides low throughput
- Requires an environmental chamber for controlling Temperature, humidity and reducing noise

There are two types of optical Lithography: DUV (Deep Ultra Violet Light) and EUV (Extreme Ultra Violet Light) which are discussed in the following subsections.

### 5.3.1 Deep Ultraviolet Lithography

Deep Ultraviolet Lithography reduces minimum feature size of the integrated circuits by 50nm. 200-300nm replaces wavelengths of 250-350nm. We use excimer light sources (used in production of microelectronic devices) to generate DUV light [4]. Currently 248nm (KrF) and 193nm (ArF) excimer lasers [6] are used for development and volume manufacturing of ICs. It is also used for commercial purposes where xenon mercury lamps with low intensity are used [7].

#### 5.3.1.1 Advantages:

- It helps in saving time and cost of manufacturing ICs is less.
- It helps in manufacturing large amount of chips at one time. We can fabricate 200 or 1000 chips instead of one easily.
- Wide ranges of resists are available.

#### 5.3.1.2. Disadvantages:

- Since the wavelength of light is getting smaller, glasses lenses designed to focus the light rather absorb the light and very less light is able to reach on the silicon wafer.
- Lack of availability of lamps.
- A mismatch can occur between the lamp and the resist used.

### 5.3.2 Extreme Ultraviolet Lithography

In order to keep pace with growing demands of manufacturing small feature size integrated chips, a new way of reducing the wavelength has been discovered using Extreme Ultraviolet Lithography. Deep ultra violet Lithography used radiation of wavelength of 248 nm to print the small-scale features on the wafer. Resolution = $K1 \times \lambda \div NA$

Where $\lambda$ is the wavelength and NA is the numerical aperture. K1 here is 0.25, since below that double patterning has to be applied which is costly and numerical aperture should not exceed 1.35 as it requires high index lenses and fluid which are not currently available.
Resolution = $0.25 \times 193 \div 1.35$ = 36nm

One of the important Problems is reducing the wavelength of light and extreme ultraviolet Lithography comes into the picture and plays an important role in optical Lithography.

#### 5.3.2.1. Procedure

*Step 1:* Wavelength of 13.5nm is used for EUV. In order to produce such small wavelength, an excited state atom is required and therefore tin (Sn) is commonly used which is already highly positively charged. Vacuum plays an important role since wavelength of light is so small that it can easily be absorbed by air. Accordingly everything ranging from optics to mask is placed inside a vacuum to generate plasma.

*Step 2:* There are two methods through which high-energy photons can be generated for making the pattern on the wafer [11]. Figure 6 shows the layout for producing EUV Lithography.

a. Laser Pulsed Plasma:
   i. A pulse laser is fired onto the incoming tin (Sn) droplets which puts these droplets in the excited state and when comes to relaxed state generates EUV photons. A mirror focuses these photons to an intermediate focus (If) and passes it to the next level. The size of the tiny tin (Sn) droplets is usually of diameter 50-100μm.

b. Discharged Pulsed Plasma:
   i. Plasma can also be produced by Discharged rather than laser pulsed plasma where two tin baths are used which are filled by molten tin and two disks are placed which rotate and pick up tin on their surface when rotated in tin bath. Apply voltage and throw laser which puts the tin to an excited state and when it comes to relaxed state it generates EUV photons.

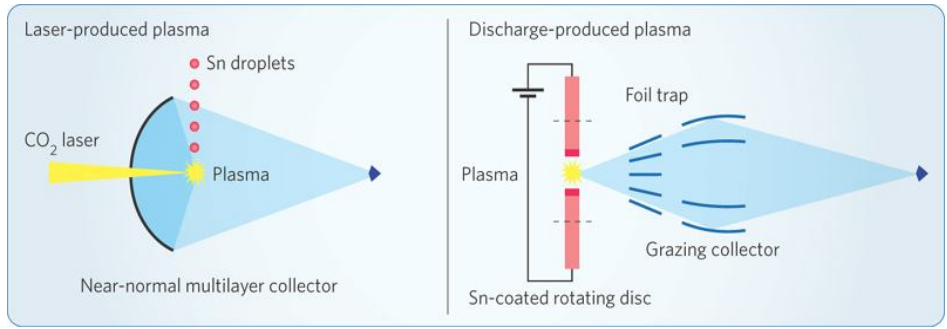

Figure 6. Layout for producing EUV Lithography

*Step 3:* After the photons get generated, concave and convex mirrors and not glasses are used since it does not refract instead reflect.

*Step 4:* A mask, which contains the pattern, is acted upon by light, which is focused onto the particular pattern while some light passes through it without even touching the pattern. EUV Lithography involves reflective optics as well as reflective mask and Brag reflector is used to make this happen.

*Step 5:* Brag Reflector: Each of the mirrors contain alternate layers of molybdenum (Mo) and silicon layers which give a reflection of about 70 percent and together form a brag reflector and will only reflect the light coming at 13.5nm and others will just pass away.

*Step 6:* The light then makes the pattern of the mask on the EUV resist coated wafer.

**5.3.2.2. Advantages:**

- Helps in producing minimum feature size.
- Numbers of transistors have increased successively which have resulted in greater speed.
- Provides high throughput.
- Provides good resolution due to low value of wavelength.

**5.3.2.3. Disadvantages**

- *Source Problems*
  1. Brightness: The laser producing plasma is too dim .It has a factor of 10-100x and less throughput is obtained
  2. Reliability and time problems are commonly faced. EUV process can't be carried out the whole day since source available is only 0.7 percent and the collector used for tin gets damaged.
  3. Tin debris: sometimes laser doesn't hit the tin droplets and get collected to form debris on the collector which can damage the collector and needs to be removed.

- *Mask Problems*
  1. Any defects on the mask that gets printed on the wafer will destroy the wafer and defect free mask is required but for EUV defect free mask is not available.
  2. Adequate technology is not available which can easily detect any kind of defect with printing nor we have repair methods, which can easily repair the defects, and also if there is defect in the molten layer, then there is no method to repair that defect.

  3. In reality only $0.3814^{th}$ of photons actually make it to the wafer since we know that ranging from source through the mask onto the wafer the reflectivity is 0.7 percent.

- *Resist*
  1. Resists used are very similar to those used in wavelengths 248 and 193 nm resists
  2. All previous resists had photons hitting at them with energy 6 or 7ev but now the energy is 92ev, so with such high energy hitting the resist, it generates secondary electrons and photon electrons which lead to degradation of image quality.
  3. Line width roughness and sensitivity are inversely proportional. Higher the sensitivity lower will be the roughness, which means much higher dose of photons is required to form the patterns, which depict low roughness.

## 6. CONCLUSION

Lithography is the base to excel in in the semiconductor industry. For many applications smaller is better in terms of speed, power consumption and cost. This paper discussed future lithographic techniques briefly. While the critical dimension in the microprocessor industry is constantly going down; stringent demands should be made on contamination control, yield enhancement and particle detection. There still remains debate over costs of many of these methods and the ability to manufacture integrated circuits in large volume. The ability to manufacture small devices is crucial for applications in many different areas including microwave circuits, high-speed optics as well as bio-electronic devices. There is a demand for high-resolution Lithography capability combined with extreme reproducibility over large areas. Historically photolithography has been the dominant technique for replication since the origin of microelectronic devices. Several other techniques are on the edge of technological breakthroughs.

### Author

**Mr. Jatin Chopra** is currently pursuing B.E (Hons) from Birla Institute of Technology & Science (BITS), Pilani in the department of Computer Science in Dubai, UAE.

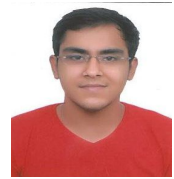